\documentclass{aa}
\usepackage{graphics}
\begin{document}
\title{Line profile variations in classical Cepheids.\\
 Evidence for non-radial pulsations?
\thanks{Based on spectra collected at McDonald 2.1m and Kitt Peak 4m (USA),
 CTIO 4m (Chile) and MSO 1.8m (Australia)}}
\titlerunning{Line profile variations of Cepheids}
\author{V.V. Kovtyukh  \inst{1,2},\, S.M. Andrievsky \inst{1,2},\, R.E.
Luck \inst{3}, and N.I. Gorlova \inst{1,2,4}}
\authorrunning{Kovtyukh et al.}
\offprints{V.V. Kovtyukh}
\institute{
Department of Astronomy and Astronomical Observatory of Odessa
State University, Shevchenko Park, 65014, Odessa, Ukraine
\and
Isaac Newton Institute of Chile, Odessa Branch, Ukraine\\
email: val@deneb.odessa.ua; scan@deneb.odessa.ua
\and
Department of Astronomy, Case Western Reserve University, 10900 Euclid
Avenue, Cleveland, OH 44106-7215\\
email: luck@fafnir.astr.cwru.edu
\and
Department of Astronomy, The University of Arizona,
Tucson, AZ, USA 85721
}
\date{Received 30 July 2002; accepted 14 January 2003}
\abstract {We have investigated line profiles in a large sample of Cepheid
spectra, and found four stars that show unusual (for Cepheids) line profile
structure (bumps or/and asymmetries). The profiles can be phase dependent but
the behavior
persists over many cycles. The asymmetries are unlikely to be due to the
spectroscopic binarity of these stars or the specific velocity field in their
atmospheres caused by shock waves. As a preliminary hypothesis, we
suggest that the observed features on the line profiles in the spectra of
X Sgr, V1334 Cyg, EV Sct and BG Cru can be caused by the non-radial
oscillations. It is possible that these non-radial oscillations are connected
to resonances between the radial modes
(3.$^{\rm d}$2, 7.$^{\rm d}$5 or 10.$^{\rm d}0$).
}

\maketitle

\section{Introduction}

The problem of line profile variation in classical Cepheids has been
a subject of many studies, for example, Karp (\cite{karp75a}; \cite{karp75b}),
Butler, Bell \& Hindsley (\cite{bbh96}).
Generally speaking, the profile variations reflect the
movements of the gas in radially pulsating atmosphere, and nowadays they
are well understood in the framework of radiative hydrodynamics. The shifts
of the line center mass and time-dependent profile asymmetry are
characteristic features of all classical Cepheids. Along with this, some
Cepheids were reported to demonstrate an unusual behaviour of their line
profiles. Evidence of a strange doubling of a number of low-excitation
lines in a classical Cepheid was firstly reported by Kraft (\cite{kraft56})
for X Cyg (P=16.$^{\rm d}$386). However, subsequent observations of this star
by Butler (\cite{butler93}) and our data (20 spectra, unpublished) did not
recover this effect. Kraft (\cite{kraft67}), though, has shown that
line doubling in X Cyg varies from cycle to cycle.

Later, Sasselov, Lester \& Fieldus (\cite{slf89}), and then Sasselov \&
Lester (\cite{sl90}) investigated unusual structure in the line profiles
of the Cepheid
X Sgr. They noticed line doubling (splitting) in X Sgr using infrared spectra.
Figures from those papers clearly show the additional blue- or red-shifted
absorptions in the line profiles. For instance, both blue and red absorptional
components together with a central absorption are clearly seen at $\phi$(dyn)
= 0.13 (see Fig. 12 in Sasselov \& Lester \cite{sl90}). The authors also
pointed out the stability of this phenomenon during the 11 month period
spanned by their observations. In 1990, Sasselov \& Lester made an attempt to
interpret the phenomenon described by Sasselov, Lester \& Fieldus (1989) as
a result of an entangled atmospheric velocity field, specifically by supposing
the existence of pulsationally-driven shock-waves and radiative transfer
along their path in the atmosphere. The main difficulty of such a rather loose
interpretation is that while the additional absorption components at the line
profiles are present over the whole period of pulsation, it is impossible for
the shock waves to propagate continuously during the whole cycle.

In 1999, Kovtyukh \& Andrievsky detected anomalous line splitting
(i.e. satellite
absorption at the profile of each line) in the spectrum of another classical
Cepheid, EV Sct, and interpreted it as a sign of spectroscopic binarity of
this Cepheid.
Kiss \& Vink\'o (\cite{kv00}) discovered additional absorption components in
the spectral lines of a classical Cepheid V1334 Cyg, and also
attributed this phenomenon to a fourth component in the system of this Cepheid.
However, such an interpretation of line splitting based on possible
binarity suffers from some weak points (see Sect. 3.2), and apparently another
hypothesis is required to explain such a phenomenon. For this, first of all,
it is necessary to find the stars among Cepheids where such a phenomenon is
seen, and then to investigate their properties.

Having at our disposal a large data-base of Cepheid high-resolution spectra
(multiphase observations) collected by us during the past few years,
we have searched for visible manifestations of line profile anomalities
in these stars. In total, more than 700 spectra of 99 galactic Cepheids
were analysed. Part of these spectra were used in galactic
metallicity gradient studies (Andrievsky et al. \cite{andet02}, and references
therein), and a description
of these spectra can be found in the mentioned above papers. Among the 99
Cepheids we have found four stars, BG Cru, V1334 Cyg, EV Sct and X Sgr, with
anomalous line profile features -- additional absorption components
or unusual asymmetries (hereafter called bumps).
They are shown in Fig.1-5.

Three of them (BG Cru, V1334 Cyg, EV Sct) are s-Cepheids (it is quite
possible that X Sgr is also an s-Cepheid, see Sect.3.4 and Fig.7).
The so-called s-Cepheids (Cepheids with $sinusoidal$ light curves and small
amplitudes) are first overtone pulsators. They were discriminated from
other classical Cepheids first by qualitative criteria, and then by
a more precise quantitative definition, based on the Fourier decomposition
of the light curves, was introduced by Antonello, Poretti \& Reduzzi
(\cite{antet90}). Microlensing surveys (MACHO and EROS) have unambiguously
shown that $all$ s-Cepheids pulsate in the first (or second) overtone
(Welch et al. \cite{welchet95}; Beaulieu et al. \cite{beaulet95}).

\begin{figure}
\resizebox{\hsize}{!}{\includegraphics{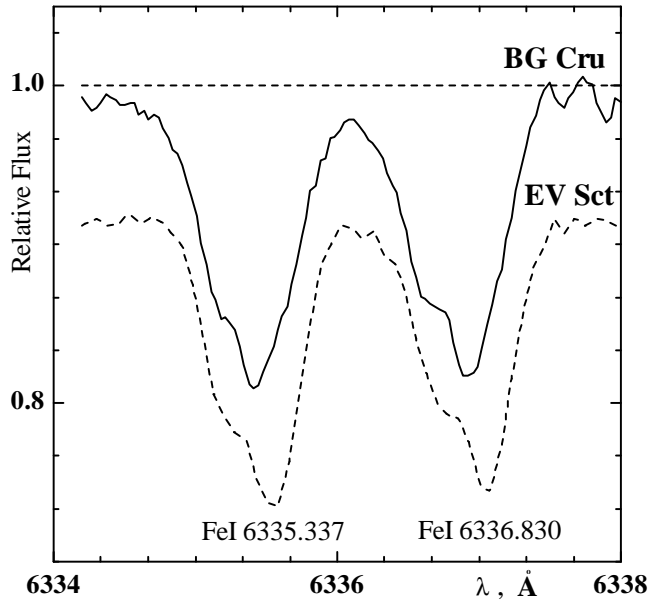}}
\caption[]{Absorption bumps in the spectra of BG Cru ($\phi$=0.28) and EV Sct
($\phi$=0.65). The EV Sct spectrum is vertically shifted.}
\end{figure}

The main characteristics of BG Cru, V1334 Cyg, EV Sct and X Sgr
are given in Table 1. In this table we also list some properties of a
number of other Cepheids that will be discussed in Sect. 3.3.

\begin{table*}
\caption[]{Program stars. Cepheids with bumps: upper panel, and\\
additional Cepheids with high FWHM values: lower panel}
\begin{tabular}{rlrrrrr}
\hline
 Star & Type& P, day & $<$V$>$ & (B-V) & A$_{V}$ & FWHM, \AA\\
\hline
\multicolumn{7}{c}{Cepheids with bumps:}\\
\object{ BG Cru }& SCEP & 3.3427&  5.487&      0.606&    0.246&  0.735\\
\object{V1334Cyg}& SCEP & 3.3330&  5.871&      0.504&    0.146&  0.639\\
\object{ EV Sct }& SCEP & 3.0909& 10.137&      1.160&    0.300&  0.67:\\
\object{ X Sgr  }& DCEP:& 7.0129&  4.549&      0.739&    0.590&  0.669\\
         &     &       &       &           &         &       \\
\multicolumn{7}{c}{Other Cepheids with large FWHM values:}\\
\object{ FM Aql }& DCEP& 6.1142&  8.270&      1.277&    0.724&  0.647\\
\object{ RX Aur }& DCEP&11.6235&  7.655&      1.009&    0.664&  0.614\\
\object{ DL Cas }& DCEP& 8.0007&  8.969&      1.154&    0.571&  0.644\\
\object{ SU Cyg }& DCEP& 3.8455&  6.859&      0.575&    0.766&  0.615\\
\object{ Y Lac  }& DCEP& 4.3238&  9.146&      0.731&    0.705&  0.622\\
\object{V340 Nor}& SCEP&11.2870&  8.375&      1.149&    0.310&  0.653\\
\object{ Y Sgr  }& DCEP& 5.7734&  5.744&      0.856&    0.725&  0.609\\
\object{ S Vul  }& DCEP& 68.464&  8.962&      1.892&    0.588&  0.601\\
\hline
\end{tabular}
\end{table*}

\section{Possible origin of the detected peculiarity}

As one of the possible origins of the line profile anomalies in the mentioned
above Cepheids, non-radial oscillations could be considered.

For years it was believed that non-radial pulsations, being a primary
attribute of B--A--F main sequence stars like $\delta$ Sct or $\beta$ Cep
pulsating variables, are at the same time not seen (i.e. not excited) in
pulsating F--G supergiants (Cepheids, for example). Dziembowski (\cite{dziem71})
argued that $p$-wave modes in stars with steep mass concentration
towards the center, i.e. in (super)giants, cannot exist because of
significant energy dissipation. Osaki (\cite{osaki77}) was the first to
show that
under special conditions, F--G supergiants can be unstable against non-radial
oscillations. He considered the atmosphere of a pulsating star as an isolated
oscillating zone with a progressive-wave boundary condition at its bottom,
and concluded that Cepheid-like stars should be vibrationally
unstable not only for radial pulsations, but also for non-radial modes with
high spherical harmonic numbers. Osaki (1977) left the question of the
actual existence of non-radial pulsation in supergiants to future
studies, since up to then no observational evidence of such oscillations
was available for Cepheids.

Van Hoolst, Dziembowski \& Kawaler (\cite{hoolstet98}) considered the
possibility of the
excitation of non-radial pulsations in classical pulsating stars
(Cepheids, RR Lyrae stars, W Vir) by using the RR Lyrae model.
They found that a large number of unstable low-degree ($l$ = 1,2) modes have
frequencies in the vicinity of unstable radial mode frequencies
and proposed resonance models to explain the Blazhko effect.

The first strong evidence of the presence of non-radial modes in the
line-profile variations of a classical pulsator
was reported by Chadid et al. (\cite{chadid99}).
A detailed frequency analysis based on 669 high resolution spectra of the
Blazhko star RR Lyrae
clearly revealed the importance of non-linear effects upon the
radial fundamental mode, and a multiplet structure with a separation equal
to the Blazhko frequency around the main frequency and its
harmonics.

It should be noted that classical Cepheid V473 Lyr also demonstrates an unusual
amplitude modulation of the light and radial velocity on a time-scale of
about 1000 days. Van Hoolst \& Waelkens (\cite{vhw95}) have interpreted such
behaviour
as a resonant interaction between the second overtone and a non-radial mode
with approximately the same period (see also Koen \cite{koen01}).

Another illustrative example among the supergiants is Polaris ($\alpha$ UMi),
a low-amplitude s-Cepheid with a period of 4.0 days. Hatzes \& Cochran
(\cite{hc00}) have found a residual component in high-accuracy radial
velocities with an amplitude of 400 m s$^{-1}$ and a period of about 40 days.
They argue that this is more likely to be due to a non-radial pulsation,
rather than to a low-mass companion or to rotational modulation from spots.
Non-radial $g$-modes are probably present in the F8Ia supergiant V810 Cen
(Kienzle et al. \cite{kienet98}), and in yellow hypergiants like
$\rho$ Cas (Lobel et al. \cite{lobelet94}).

Butler (\cite{butler98}) presented precision velocity results from a 6 yr
survey of 15 supergiants that lie in (or near) the Cepheid instability strip.
Periodograms of many of these stars show significant peaks at 50-80 days
which are unlikely to be associated with radial pulsation.

\section{Discussion}

\subsection{The bumps}
     
As it is known, the well-studied non-radial pulsators among B, A-stars
show either bumps (they are the common feature of fast rotators, see,
for example, Uytterhoeven et al. \cite{uytt01} for $\kappa$ Sco,
Gies \& Kullavanijaya \cite{gk88} for $\epsilon$ Per), or some kind of line
asymmetry in the slow rotators (Aerts et al. \cite{aerts94} for
$\beta$ Cep stars, Aerts et al. \cite{aerts99} for Slowly Pulsating B stars).

Figs. 1-5 leave no doubt that absorptional peculiarity of the line profiles
(bumps) really exists in the spectra of four Cepheids listed in the upper panel
of Table 1. For EV Sct, only the blue bump is clearly seen, while both
blue and red bumps are visible in the other three Cepheids. This could be due
to the small number of EV Sct spectra which were perhaps observed at phases
that are not appropriate for simultaneous detection of two bumps. One should
note that the bumps are seen on the profiles of all unblended spectral lines.
The most prominent bump features and time evolution were detected in X Sgr.

Using the blend separation algorithm of Cassatella (\cite{cas76}), we
deconvolved the X Sgr, V1334 Cyg, BG Cru and EV Sct line profiles in order
to find the blue/red bump velocity shift relative to a central absorption.
For BG Cru and EV Sct we obtained the following $\Delta$V$_{\rm r}$ values
for the blue bump: $-16.43 \pm 0.19$ km~s$^{-1}$ for BG Cru and
$-16.88 \pm 0.14$ km~s$^{-1}$ (1992), and $-16.40 \pm 0.48$ km~s$^{-1}$
(2002) for EV Sct.

A sufficient number of spectra of V1334 Cyg and X Sgr, and pronounced
bumps, allowed us to resolve phase-dependent changes of the radial velocity
of all absorption components in the line profiles. For example, in the case
of V1334 Cyg, one can see the flat line bottom as a sum of two equal
absorption bumps at $\phi$=0.365, the blue bump at $\phi$=0.660 and the
red bumps at $\phi$=0.819, 0.960, 0.997 (Fig. 3). The radial velocity
separation between the displaced absorption bumps and the central absorption
component is +15.1 km~s$^{-1}$ and --12.4 km~s$^{-1}$ at phases 0.819 and
0.365 respectively, but with a large uncertainty of about
$\pm 3-4$ km~s$^{-1}$.

\begin{figure}
\resizebox{\hsize}{!}{\includegraphics{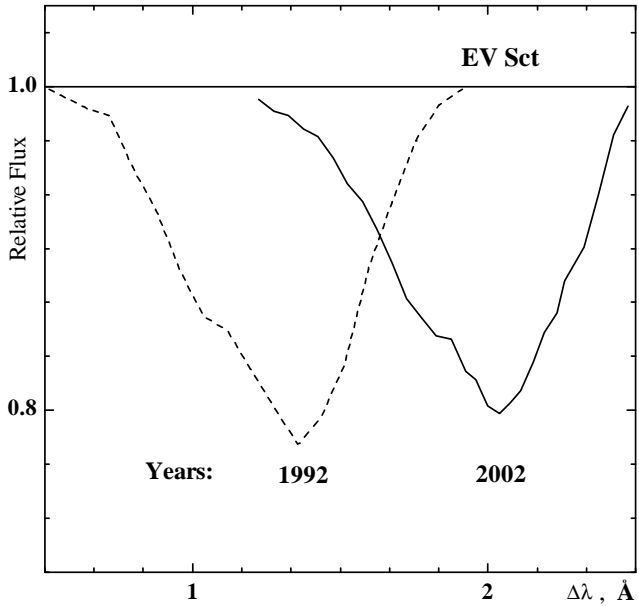}}
\caption[]{Averaged line profiles of EV Sct in 1992 and 2002 years.
Each profile is a sum of approximately twenty unblended line profiles.}
\end{figure}

\begin{figure}
\resizebox{\hsize}{!}{\includegraphics{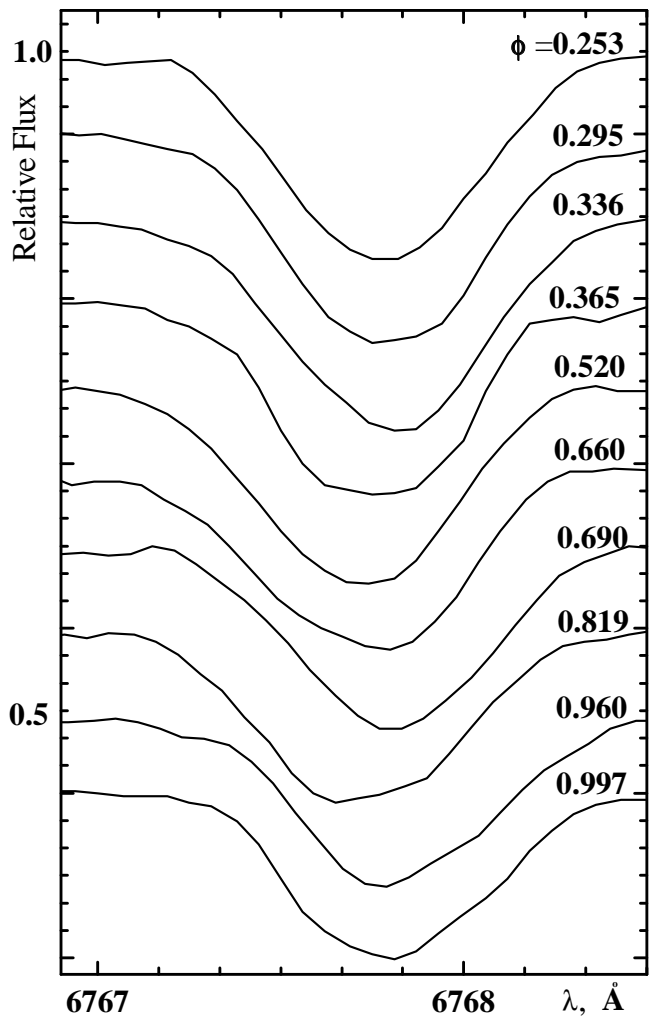}}
\caption[]{Observed Ni I 6767.78 \AA \, line profiles in V1334 Cyg spectrum.
Phases are indicated near the profiles.}
\end{figure}

In general, the character of the profile change (transition from blue bump to
red one) in V1334 Cyg resembles very much that in the X Sgr spectrum analyzed
in this paper (see Fig. 4-5). We have also re-examined the spectra of X Sgr
obtained by Luck \& Lambert (\cite{ll81}, \cite{ll85}) in 1978-1979, and find
absorption bumps (note that Luck \& Lambert did not consider the
multicomponent structure of the line profiles). Thus, one can state that the
discussed phenomenon has been persistent for 20 years. Comparison of line
profiles from the 1978 and 1997 spectra at similar phases shows an increase
of the blue component strength and weakening of the red one. However, this is
not a very confident result as the spectra have different resolutions.

The radial velocity measurements carried out by us, similar to those Sasselov
\& Lester (\cite{sl90}), for both satellite absorptions are shown in Fig. 6.
The corresponding phases are calculated according
to the quadratic elements from Berdnikov \& Ignatova (\cite{bi00}). From
Fig. 6 one can conclude that the absorptions are quasi-equidistant
in radial velocity space (within the errors of radial velocity measurements
which are, nevertheless, rather high). The two satellite absorption features
are always shifted at a constant value with respect to the main absorption
feature: $+21.3 \pm 1.8$ km~s$^{-1}$ for the red one and $-24.8 \pm
4.0$ km~s$^{-1}$ for the blue one, which agree with each other to within
the errors of measurement.

\begin{figure}
\resizebox{\hsize}{!}{\includegraphics{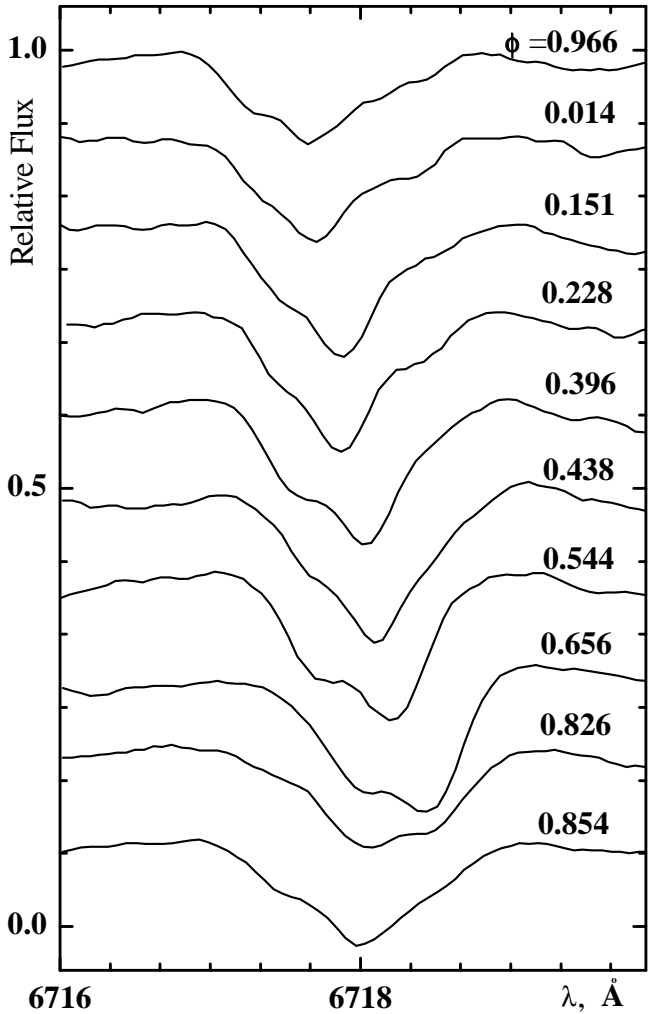}}
\caption[]{ A sample series of  Ca I 6717.69 \AA \, line profiles of X Sgr
showing the evolution of the blue to red moving bump structure.}
\end{figure}

\begin{figure}
\resizebox{\hsize}{!}{\includegraphics{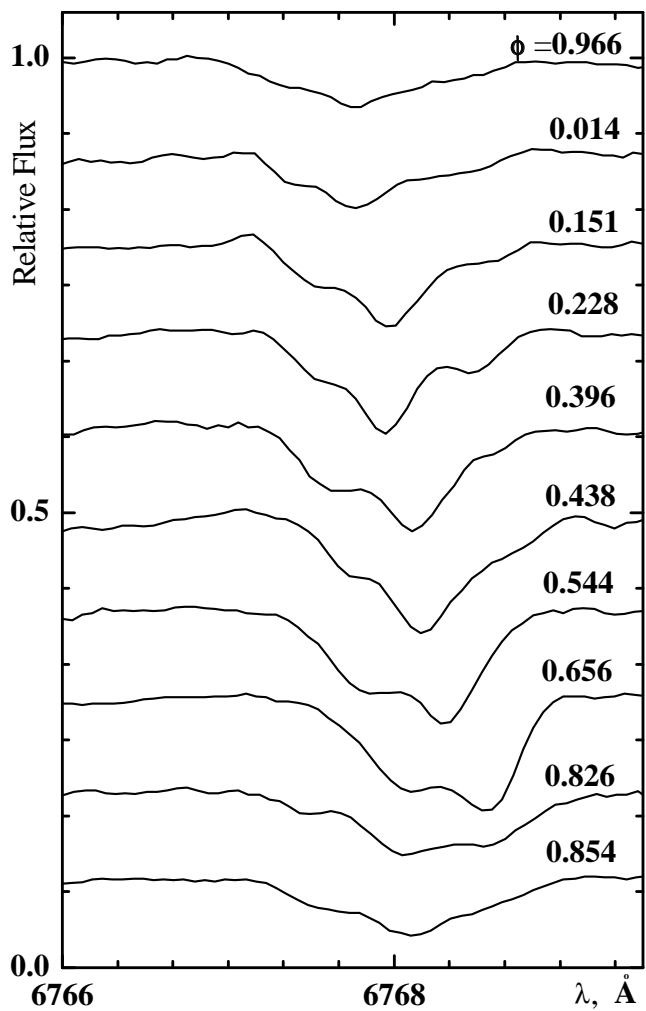}}
\caption[]{Same as in Fig.4 but for Ni I 6767.78 \AA \, line.}
\end{figure}

The line profile bumps in our four Cepheids resemble those that are
seen in line profile of non-radially pulsating stars
(e.g. Osaki \cite{osaki71}; Smith \cite{smith80};
Gies \& Kullavanijaya \cite{gk88}).
According to Shibahashi \& Osaki (\cite{so81}), the harmonic index $l$ shifts
to higher values with increasing T$_{\rm eff}$. In connection with this,
it is interesting to note that our four candidates for non-radial pulsators
lie well in the vicinity of the blue edge of the instability strip, as
illustrated by Fig. 7. To construct this diagram, the phase-averaged
effective temperatures of all our stars were determined following
Kovtyukh \& Gorlova (\cite{kg00}). Absolute visual magnitudes were obtained
with the help of the "period-luminosity" relation of Gieren, Fouqu\'e
\& G\'omez (\cite{gfg98}).

\subsection{Non-radial pulsation vs. binarity}

In principle, the line splitting of a spectroscopic binary at some phases
can mimic a bump feature resulting from non-radial pulsations, and
$vice~versa$. If blue and/or red absorption features are detected, but
long-term spectroscopic observations are not available, then it is difficult
to positively conclude whether one is dealing with a spectroscopic
binary star or a non-radial pulsator.

Therefore, it is quite important to search for some independent evidence
of the possible spectral binarity of our four Cepheids suspected to be
simultaneously the radial and non-radial pulsators.

\begin{figure}
\resizebox{\hsize}{!}{\includegraphics{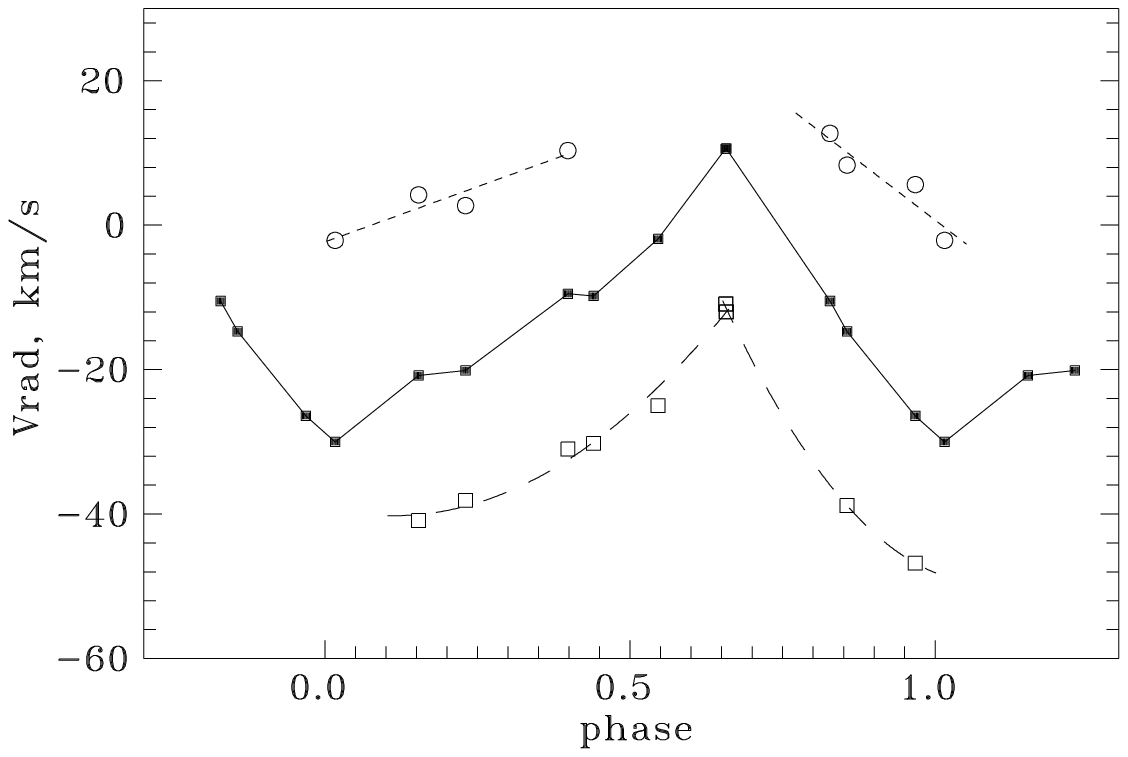}}
\caption[]{Radial velocity curves for X Sgr. The filled squares are the
velocities of the main absorption component of the line. The bump velocities
are shown as open squares (blue one) and open circles (red one).}
\end{figure}

Berdnikov and Pastukhova (\cite{bepa95}) detected $O-C$ variations for
EV Sct with a probable period of $\approx$10000 days. A large scatter in
the light and radial velocity curves has been noted in the work of Pel
(\cite{pel76}) and Mermilliod et al. (\cite{meret87}). This could be a sign
of EV Sct duplicity. As it was mentioned in the Introduction, Kovtyukh \&
Andrievsky detected line splitting in the spectrum of EV Sct, and attributed
it to spectroscopic binarity of this Cepheid. The most recent spectral
observations of EV Sct, described in the present paper, confirm an existence
of the blue bump (Fig. 2). Moreover, both spectra, separated by 10 years but
taken at similar pulsation phases, show practically the same velocity
difference between the bump and central absorption ($-16.88$ km~s$^{-1}$ and
$-16.40$ km~s$^{-1}$ respectively). This does not favor the hypothesis of
spectroscopic binarity, as otherwise, one would expect to see a different
velocity between the line components (unless the binary period is very close
to 10 years).

Kiss and Vinko (\cite{kv00}) described the line splitting in V1334 Cyg spectrum,
and interpreted such a splitting as a binarity sign. According to Henriksson
(\cite{hen82})  V1334 Cyg was known before as a triple star containing an
anomalously low-amplitude Cepheid (A$_{\rm V}$ = 0.$^{m}$146), a secondary
of B5--B8 spectral class (V=7.$^{m}$9, at a distance of 0.13 arcsec for
epoch 1975.0), and another fainter companion that was detected from
the Cepheid's velocity variations on a time-scale of about 5 years
(Evans \cite{evans00}). Nevertheless, both known companions of V1334 Cyg
are too faint to contribute significantly to the optical spectrum.

BG Cru is a poorly studied s-Cepheid with a small amplitude (A$_{\rm V}$ =
0.$^{\rm m}$203). Duplicity was suspected on the basis of the low amplitude
light variations in the U band (Dean \cite{dean81}). Szabados (\cite{szab89})
reached
the same conclusion using V$_{\gamma}$ variations. Evans (\cite{evans92})
failed to detect a hot companion with IUE data and assigned an upper limit
of A1 for the spectral type of the undetected companion.

Evans (\cite{evans92}) provided a lower limit of spectral type A0 for
the imputed companion of X Sgr.

Summarizing, the bumps observed in X Sgr, V1334 Cyg, EV Sct and BG Cru spectra
are unlikely to be to duplicity (or multiplicity) of these Cepheids. Especially
for X Sgr and V1334 Cyg, with a large number of analyzed spectra, the
short-time changes of the bump positions do not
support the companion-based hypothesis. For BG Cru and EV Sct additional
observations are warranted in order to further discriminate between the two
possibilities.

\subsection{Other peculiarities of four program Cepheids}

Another interesting feature of our four Cepheids are anomalously broad
spectral lines (for V1334 Cyg this was also found by Kiss \& Vink\'o
\cite{kv00}, and for EV Sct this was noted by Bersier \& Burki \cite{bb96}).
One should mention that the line FWHM value in Cepheids depends upon the
phase, and reflects the effect of a global compression of the atmosphere,
as well as the shock-wave propagation (Fokin, Gillet \& Breitfellner
\cite{fgb96}; Bersier \& Burki \cite{bb96}; Kiss \& Vink\'o \cite{kv00}).
For example, Gillet et al. (\cite{gillet99}) have used FWHM to trace the
turbulent velocity variations in the well known Cepheid $\delta$ Cep. From
non-linear and non-adiabatic pulsation models they concluded that the main
factor governing the line-broadening processes is the global
compression/expansion of the atmosphere, while shock-wave effects
turned out to be much weaker.

\begin{figure}
\resizebox{\hsize}{!}{\includegraphics{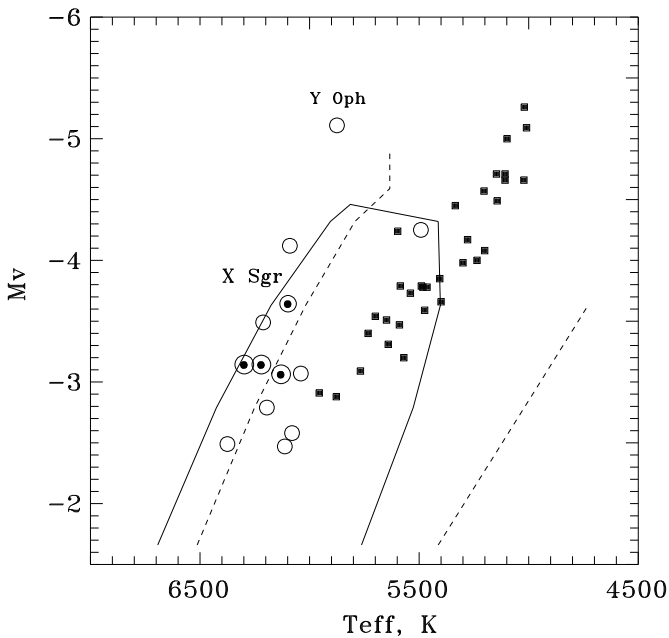}}
\caption[]{H-R diagram. The theoretical instability strip for s-Cepheids
(1-st overtone pulsators) is represented by the solid line, and for fundamental
pulsators is marked by the dashed line (turbulent convective Cepheid
models by Yecko, Koll\'ath \& Buchler \cite{ykb98}). Positions of some
observed fundamental Cepheids are shown as filled squares and s-Cepheids
as circles. The four candidates for non-radial pulsators are shown as dotted
circles.}
\end{figure}

\begin{figure}
\resizebox{\hsize}{!}{\includegraphics{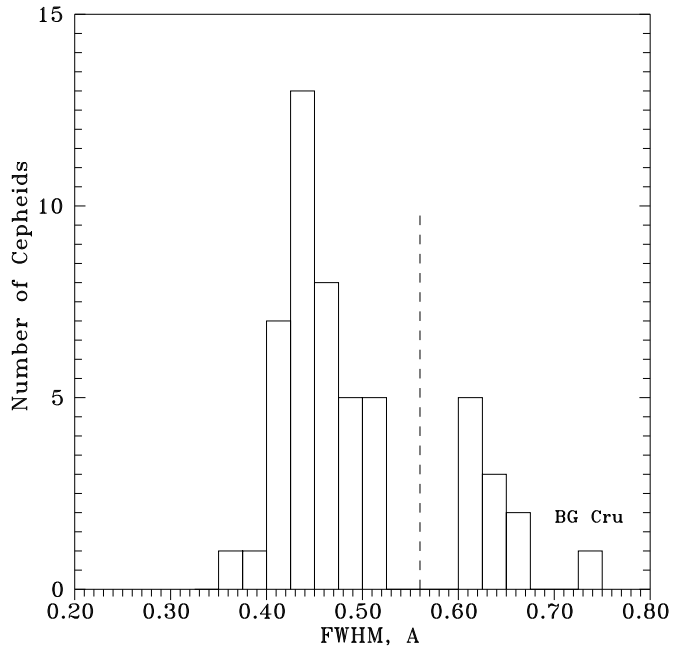}}
\caption[]{The distribution of the Cepheid line widths (FWHM, \AA) in the
phase of maximum radius is shown. Two groups are clearly separated. FWHM
for BG Cru was measured in another phase, which explains its outlying
position}
\end{figure}

For the sake of uniformity we have measured and compared the line widths
in all available Cepheid spectra at a fixed pulsation phase. The shapes of
the lines and their widths vary during the Cepheid pulsation cycle. At the
phase of maximum radius the line profiles are highly symmetric due to the
absence of systematic atmosphere motions due to pulsation. This makes the
maximum
radius phase quite suitable for line width measurements. A similar approach
was first used by Bersier \& Burki (\cite{bb96}) for 41 Cepheids (CORAVEL data).
Their conclusion was that the value of the equatorial rotational velocity of
Cepheids $v_{\rm rot} \sin i$ is smaller than 10 km s$^{-1}$, and that the
line broadening is dominated by the turbulence. Thus, they confirmed the
earlier result of Kraft (\cite{kraft66}) about very slow rotation of yellow
supergiants. The more detailed studies of Fokin, Gillet \& Breitfellner
(\cite{fgb96}) and Takeda, Kawanomoto \& Ando (\cite{tka97}) reported for
$\delta$ Cep and SU Cas $v_{\rm rot} \sin i = 7.5 \pm 2$ km s$^{-1}$ and
$v_{\rm rot}\sin i \simeq 0$ km s$^{-1}$ respectively (estimates at the phase
of near-maximum radius). It should be also noted that Bersier \& Burki
excluded from their statistics the following Cepheids: EV Sct, DL Cas and
V340 Nor as the stars having the broadest lines. The interpretation of the
broadening velocity as rotation is not the only interpretation possible.
Macroturbulence would be expected in these stars and Luck \& Lambert
(\cite{ll81}, \cite{ll85}) derive Gaussian macroturbulent velocities
averaging 8--12 km s$^{-1}$
from profile fits for a number of these stars. At the broadening velocities
determined for Cepheids it is impossible to separate a strictly Gaussian
macroturbulent profile from a rotation profile.

We have measured the Ca~I 6717.687 \AA \, line FWHM for 51 Cepheids at the
phase of maximum radius (only for those of our Cepheids the spectra near
the phase of a maximum radius are available). The determined FWHMs are
given in Table 1 (only for those Cepheids with larger than normal width
values), and the histogram of FWHMs is shown in Fig.8. While Bersier \& Burki
(\cite{bb96}) found only three stars with anomalously broad lines, we find
12 stars with $<$FWHM$>$ = $0.641 \pm 0.038$ \AA  \,
that are clearly separated
from the greater bulk of the Cepheids with $<$FWHM$>$ = $0.450 \pm 0.034$ \AA.
All four of our "bump" stars are in the broad line group.

However, the remaining 8 Cepheids from that group do not show bump features.
These Cepheids have different periods (from the 4 "bump" Cepheids) and the
only other common feature among them is a small pulsation light amplitude,
less than 0.$^{m}$8 in V, while Cepheids with narrower lines may
have larger
amplitudes (Fig.9). At present we cannot say whether this is a sign of an
anti-correlation between pulsation and line width, since for this group it
is still unclear which fraction of the line width is contributed by
rotation/macroturbulence.

BG Cru has the largest FWHM value, but this maybe simply due to the fact
that $\phi$=0.282 (only one spectrum of this star exposed at this phase
is available) is not exactly the maximum radius phase, although the lines
at this phase appear quite symmetric.

\subsection{Bumps, rotation and resonances}

X Sgr, V1334 Cyg, EV Sct and BG Cru do have broadened lines either due
to rotation or macroturbulence, and this probably substantially favors bump
detection. Nonetheless, the bumps are not seen in the spectra of the other
8 stars from the group of Cepheids with large FWHM values. What could be the
reason for this different behavior? To try to answer this question, one can
speculate about the possible connection between the non-radial mode excitation
and pulsational resonances in Cepheids.
Unfortunately, no detailed study similar to the one of Van
Hoolst et al. (\cite{hoolstet98}), based upon non-linear non-radial
oscillation theory in which the possibility of non-radial mode excitation
through resonances, exists for Cepheid models. Most current studies are
limited to resonances among radial modes.

Below we briefly summarize the known resonances for Cepheids with periods P
$\leq 10^{\rm d}$ based on the work of Moskalik, Buchler \& Marom (\cite{mbm92})
and Antonello (\cite{ant94}):

a) There is a well known resonance F$_{2}$/F$_{0}$ in Cepheids with P$_{0}$
near $10^{\rm d}$ (see, for example, Moskalik, Buchler \& Marom \cite{mbm92}).
For periods in the vicinity of $10^{\rm d}$, the primary and secondary bumps of
the light curve switch roles, and as a result the bump appears to move from
the descending to the ascending branch ($Hertzsprung$ $progression$).

b)  The resonance F$_{4}$/F$_{1}$ near P$_{1}=3.^{\rm d}2$  was suspected
by Antonello \& Poretti (\cite{antpor86}), then noted again by Petersen
(\cite{pet89}), and further described by Antonello, Poretti \& Reduzzi
(\cite{antet90}) in s-Cepheids.

c)  The resonance between the fundamental and fourth overtone mode
F$_{3}$/F$_{0}$=3 near P$_{0}=7.^{\rm d}$5 has been studied by Moskalik,
Buchler \& Marom (\cite{mbm92}) in their non-linear calculations, and
also discussed by Antonello (\cite{ant94}).

It is worth noting that our 3 s-Cepheids, i.e. the first overtone pulsators
V1334 Cyg, EV Sct and BG Cru, have their periods P$_{1}$ very close to
resonance "b" (although Kienzle et al. \cite{kienet99} on the basis of the
corresponding radial velocity data, suggest that the resonance center lies
at a much higher period, closer to 4.$^{\rm d}$6).

\begin{figure}
\resizebox{\hsize}{!}{\includegraphics{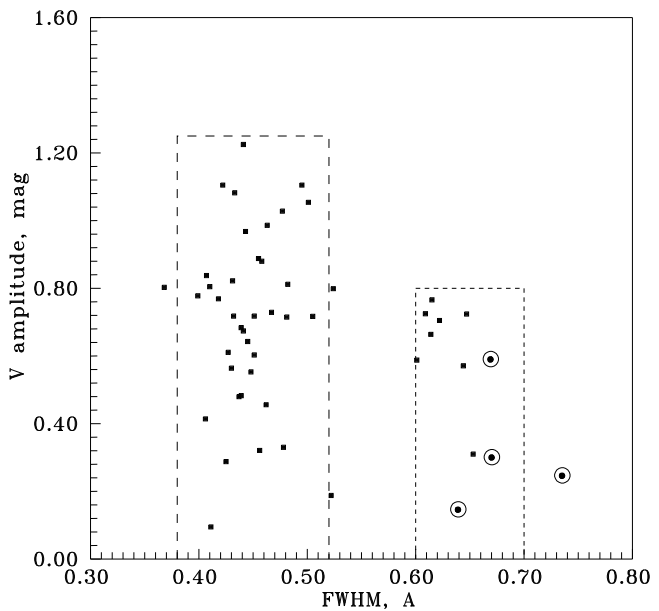}}
\caption[]{Light curve amplitudes for two groups of Cepheids with different
FWHM. One can see that the Cepheids with anomalously broad lines have lower
amplitudes  (A$_{V}<0.^{m}$8). The four candidates for non-radial pulsators
are shown as dotted circles.}
\end{figure}

X Sgr has the longest period ($7^{\rm d}.0$) of the "bump" Cepheids.
Resonance "c" could be the origin of the X Sgr bumps. On the other hand,
if we suppose that X Sgr is also a first overtone pulsator (this is
supported by its position within the HR diagram in Fig. 7), then the
period of the unexcited fundamental mode of this star would be
P$_{0}$ = P$_{1}$/0.71 $\approx 10^{\rm d}$, which fits case "a".

Interestingly, unlike them, its pulsation amplitude is not the smallest among
the Cepheids of the similar periods. We want to point out one more peculiarity
of X Sgr -- its light curve has broader maximum compared to the Cepheids of
similar periods, and a less noticeable secondary bump. We can say that the
light curve of X Sgr is intermediate between that of normal Cepheids
(fundamental pulsators) and s-Cepheids (pulsating in the 1-st overtone).
The abnormal light and radial velocity curves of this Cepheid was first
noted by Kov\'acs, Kisvars\'ayi \& Buchler (\cite{kovach90}). That is why it
is quite plausible that the star may be in a unique evolutionary stage
(switching from one pulsation mode to another, transiting to the bimodal
regime, leaving instability strip, etc), which is also corroborated by the
absence of the objects within the resonance period.

\section{Conclusion}

Four Cepheids (BG Cru, EV Sct, V1334 Cyg and X Sgr)  are found to
have unusual bump features within their spectral lines. They show
the following common peculiarities.

--  They have periods that are close to resonance values
    (around $3.^{\rm d}$2 and $7.^{\rm d}$5 or $10^{\rm d}$).

--  Their light and velocity amplitudes are the smallest among s-Cepheids
    of similar periods (V1334 Cyg, EV Sct and BG Cru).

--  The FWHM values of the spectral lines in these Cepheids are
    larger than in other ordinary Cepheids.

--  Their phase-averaged effective temperatures are rather high
    (T$_{\rm eff}> 6 000$ \,K).

As a $preliminary$ hypothesis, one can suggest that the observed bumps
in the line profiles in the spectra of X Sgr, V1334 Cyg, EV Sct and BG Cru
can be considered as a combined effect of the rather high line broadening
(either due to rotation or macroturbulence), and the
non-radial oscillations.
Taking into account that fact that characteristic periods
of those possible non-radial oscillations are quite close to observed
periods of radial pulsations in each stars, one can note that the observed
periods of the radial modes in EV Sct, BG Cru, V1334 Cyg
(and very likely in X Sgr)
are those of the first overtone, and this means that periods of the
non-radial modes are shorter than those of the (unexcited) fundamental radial
mode, as it should be, for instance, in the case of non-radial $p$-modes.
An illustrative example is given by the pulsating yellow supergiant V810 Cen.
This star has a period of the fundamental mode P$_{0} = 156^{\rm d}$,
and period of non-radial oscillations of about 107$^{\rm d}$.
This gives a period ratio of approximately 0.69
(Kienzle et al. \cite{kienet98}),
that is close to the corresponding period ratios for our four Cepheids.

Finally, we should note that the detection of the Cepheids having
bumps on their spectral line profiles is of particular interest.
The proper interpretation of this unusual phenomenon can help
1) to verify the existing pulsational models of Cepheids (this is
quite important for the determination of accurate Cepheid masses),
and 2) to explain the existence of Cepheids with broadened line profiles,
that has no explanation at present.

\begin{acknowledgements}
The authors would like to acknowledge Drs. A. Fry, B. Carney, D. Bersier,
M.R. Meyer and E. Mamajek for their help obtaining spectral material,
Drs. V. Gopka and A. Yushchenko for their version of the Cassatella
blend-separation code, and also to Dr. V. Pariev for a very helpful
discussion.
The authors are indebted to Dr. L. Balona for his very valuable comments,
and to the anonymous referee for careful reading of our manuscript and
numerous important remarks that helped to improve the paper.
\end{acknowledgements}

\end{document}